\documentclass[12pt]{article}
\usepackage[totalheight = 23cm, totalwidth = 17cm]{geometry}

\newcommand{\be}{\begin{equation}}
\newcommand{\ee}{\end{equation}}
\newcommand{\bea}{\begin{eqnarray}}
\newcommand{\eea}{\end{eqnarray}}

\newcommand{\nn}{\nonumber\\}

\newcommand{\ol}{\overline}
\newcommand{\X}{{\bf X}}

\newcommand{\z}{{\bf z}}
\newcommand{\D}{{\cal D}}

\usepackage{graphicx}

\begin{document}

\begin{center}

{\bf{\Large Non critical superstring configuration and Minkowski space time in four dimensions} 

\vspace{.3cm}

Jean Alexandre} 

\vspace{.2cm}

{\it Department of Physics, King's College London,

The Strand, London WC2R 2LS, U.K.}

\vspace{2cm}

{\bf Abstract}

\end{center}

We extend the non-perturbative time-dependent bosonic string action
of \cite{AEM1} to a $N=1$ supersymmetric world sheet action with graviton background,
and assume a superpotential, function of the time super coordinate.

\vspace{2cm}

The use of (super)strings for the description of cosmological events suggests that 
String Theory might become observable, given the huge amount of cosmological data
\cite{danielsson}. 
String Cosmology is generally based on the study of the effective theory in the target space
\cite{tseytlin}, where the effective action is built in such a way that the corresponding equations
of motion give the perturbative Weyl invariance conditions on the world sheet of the string.
Instead, in \cite{AEM1}, a time-dependent bosonic string configuration was found to satisfy Weyl
invariance conditions in a non-perturbative way: it was conjectured that this configuration cancels the ressumation of
all loops in the beta functions corresponding to Weyl invariance. The corresponding action is given by
\be\label{bosonic}
S=\frac{1}{4\pi\alpha^{'}}\int d^2\xi\sqrt{\gamma}\left\{\gamma^{ab}\frac{\kappa_0\eta_{\mu\nu}}{(X^0)^2}
\partial_a X^\mu\partial_b X^\nu+\alpha^{'}R^{(2)}\phi_0\ln(X^0)\right\},
\ee
where $\kappa_0$ and $\phi_0$ are constants. 
Weyl invariance is then satisfied for any target space dimension $D$ and any dilaton amplitude $\phi_0$.
The metric amplitude $\kappa_0$ is then a function of $D$ and $\phi_0$, which is not known precisely, but which
has no cosmological relevance.

The string configuration (\ref{bosonic}) describes of a spatially-flat 
Robertson Walker Universe, with metric
$ds^2=-dt^2+a^2(t)(d\vec x)^2$, where $t$ is the time in the Einstein frame, and $a(t)$ is the scale factor.
For the configuration (\ref{bosonic}), the scale factor was shown to be a power law
\be
a(t)\propto t^{1+\frac{D-2}{2\phi_0}},
\ee
such that, if the following relation holds
\be\label{Minkowski}
D=2-2\phi_0,
\ee
the target space is static and flat (Minkowski Universe).

We are interested here in a superstring action with corresponding bosonic part (\ref{bosonic}),
and we consider for this a $N=1$ world sheet supersymmetric action, with graviton background, and in 
which we assume a superpotential $\Phi$, function of $\X^0$, where $\X^\mu$ are the supercoordinates
on the world sheet.

\vspace{0.5cm}

The string supercoordinates of an NSR string are \cite{polchinski}
\be
\X^\mu=X^\mu+i\theta\psi^\mu+i\ol\theta\ol\psi^\mu+\theta\ol\theta F^\mu,
\ee
where $\psi^\mu$ and $\ol\psi^\mu$ are complex conjugate if the world sheet is Euclidean.
The supercovariant derivatives are ${\cal D}=\partial_\theta+\theta\partial_z$ and
$\ol{\cal D}=\partial_{\ol\theta}+\ol\theta\partial_{\ol z}$, where 
$\z=(z,\theta)$ denotes the superspace coordinates on the world sheet.
The $N=1$ supersymmetric world sheet action we consider is
\be\label{action}
S=\frac{1}{4\pi\alpha^{'}}\int d^4\z\left\{\kappa(\X^0)\eta_{\mu\nu}\ol\D\X^\mu\D\X^\nu+m\alpha^{'}\Phi(\X^0)\right\},
\ee
where $\kappa(\X^0)$ defines the graviton background, and $\Phi(\X^0)$ is a superpotential.
Although we consider a flat world sheet metric, where the curvature scalar $R^{(2)}$ should vanish,
we keep the latter finite and identify the corresponding scalar potential with the dilaton term:
\be
\int d\theta d\ol\theta~m\Phi(\X^0)=R^{(2)}\phi(X^0),
\ee
before taking the limit $R^{(2)}\to 0$. This procedure is proposed as a way to 
compute the superpotential giving rise to the dilaton. 

The relation between the superpotential $\Phi$ and the dilaton $\phi$ is found by expanding the 
superfields in terms of their components, and by considering the on-shell expressions for the
auxiliary fields $F^\mu$, which are
\bea
F^0&=&\frac{\alpha^{'}m}{2\kappa(X^0)}\Phi^{'}(X^0)-\frac{\kappa^{'}(X^0)}{2\kappa(X^0)}
\left(2\psi^0\ol\psi^0+\ol\psi^\mu\psi_\mu\right)\nn
F^j&=&\frac{\kappa^{'}(X^0)}{2\kappa(X^0)}\left(\ol\psi^0\psi^j-\psi^0\ol\psi^j\right),
\eea
where a prime denotes a derivative with respect to $X^0$.
The latter expressions generates the following potential term for $X^0$
\be\label{potentialterm}
-\frac{(\alpha^{'}m)^2}{4\kappa(X^0)}\left(\Phi^{'}(X^0)\right)^2,
\ee
such that we define $\alpha^{'}m^2=4R^{(2)}$ and
\be\label{definition}
\phi(X^0)=\frac{-1}{\kappa(X^0)}\left(\Phi^{'}(X^0)\right)^2.
\ee
In order to recover the bosonic configuration (\ref{bosonic}), $\Phi$ has to satisfy
\be
(\Phi^{'}(X^0))^2=-\kappa_0\phi_0\frac{\ln(X^0)}{(X^0)^2}.
\ee
We are interested in describing a Minkowski target space for $D=4$, where we need $\phi_0=-1$ (see eq.(\ref{Minkowski})),
and we have then
\be\label{superpotential}
\Phi(X^0)=\frac{2}{3}\sqrt{\kappa_0}\left(\ln(X^0)\right)^{3/2},
\ee
where 
\bea
\left(\ln(X^0)\right)^{3/2}&=&\left(\sqrt{\ln(X^0)}\right)^3~~~~\mbox{if}~~X^0\ge 1\nn
\left(\ln(X^0)\right)^{3/2}&=&-i\left(\sqrt{|\ln(X^0)|}\right)^3~~~~\mbox{if}~~X^0<1.
\eea
Finally, one should not make a confusion between the superpotential term in the action (\ref{action}) 
and a boundary action describing tachyon dynamics \cite{tachyon}. Instead, the superpotential (\ref{superpotential}) is
just to be seen as an intermediate step in order to take into account the dilaton of the configuration
(\ref{bosonic}). The superstring configuration leading to the bosonic action (\ref{action}) is finally given by
the action 
\be
\frac{1}{4\pi\alpha^{'}}\int d^4\z~\frac{\kappa_0\eta_{\mu\nu}}{(\X^0)^2}\ol D\X^\mu D\X^\nu,
\ee
together with the dilaton appearing in (\ref{bosonic}).

\end{document}